\documentclass[hyper]{JHEP3}

\usepackage{amsmath}
\usepackage{amscd}
\usepackage{amsfonts}
\usepackage{amssymb}
\usepackage{amsthm}

\renewcommand {\d}  {\partial}
\newcommand {\vd}   {\delta}
\renewcommand {\phi}{\varphi}
\renewcommand {\a}  {\alpha}
\newcommand {\N} {\mathcal{N}}
\renewcommand {\l}  {\lambda}
\newcommand {\s}  {\sigma}

\newcommand {\ez} {\zeta}
\newcommand {\g}  {\mathfrak{g}}
\newcommand {\su} {\mathfrak{su}}
\newcommand {\so} {\mathfrak{so}}
\renewcommand {\sp} {\mathfrak{sp}}
\newcommand {\ev}  {\varepsilon}
\newcommand {\e}   {\epsilon}
\newcommand {\ub} {\bar{u}}
\newcommand {\Ss} {\mathcal{S}}
\newcommand {\cc} {\mathcal{C}}

\newcommand {\ind} {{ind_w}}

\newcommand {\lb} {\left (}
\newcommand {\rb} {\right )}

\renewcommand{\leq}{\leqslant}

\newcommand{\f}{\frac}

\newcommand{\tr} {\mathop{\rm Tr}}

\newcommand{\rk} {\mathop{\rm rank}}

\newcommand{\vol}{\mathop{\rm Vol}}
\newcommand{\sign}{\mathop{\rm sign}}

\newcommand{\beq}{\begin{equation}}
\newcommand{\eeq}{\end{equation}}

\newcommand{\bal}{\begin{aligned}}
\newcommand{\eal}{\end{aligned}}

\newcommand\bqa {\begin{eqnarray}}
\newcommand\eqa {\end{eqnarray}}

\title{$\mathcal{N}=4$ super Yang-Mills matrix integrals for almost
all simple gauge groups}

\author{Vasily Pestun\\
ITEP, Moscow, Russia.\\
E-mail: \email{pestun@gate.itep.ru}}

\preprint{ITEP-TH-28/02\\ \hepth{0206069}}

\abstract{
  In this paper the partition function of
  $\mathcal{N}=4$ $D=0$ super Yang-Mills matrix theory with
  arbitrary simple gauge group is discussed.
    We explicitly computed its value for all
  classical groups of rank $r \leq 11$ and for the exceptional groups $G_2$,
  $F_4$ and $E_6$. In the case of classical groups of arbitrary rank
  we conjecture general formulas for the $B_r$, $C_r$ and  $D_r$ series
  in addition to the known result for the $A_r$ series.
  Also, the relevant boundary term contributing to the Witten index of the
  corresponding supersymmetric quantum mechanics has been explicitly computed as a simple function of
rank for the orthogonal and symplectic groups $SO(2N+1)$,
$Sp(2N)$, $SO(2N)$.
}

\begin{document}

\section {Introduction}
   The super Yang-Mills  matrix models  are obtained by the dimensional reduction to $0$ dimensions
of the supersymmetric Yang-Mills  gauge theories (SYM) in several dimensions.
They have already appeared in some contexts.
Firstly, the action of $SU(N)$ model describes in the leading order the
world volume potential on a stack of $N$ $D(-1)$-branes \cite{Witten:1995im}.
Secondly, these models (in the limit $N \to \infty$) are
relevant for the constructions \cite{Banks:1996vh} which are believed
to provide a non-perturbative formulation of superstring theory.
Since the dynamic of the reduced theory can capture
some features of dynamic of the full unreduced theory, it is not surprising
to encounter them in multi-instanton calculus \cite{Dorey:1998qh}.
Finally, they are closely connected with the question about the number of
normalizable ground state in the SYM quantum mechanics \cite{Porrati:1997ej,Yi:1997eg,Sm2,Sethi,GrGut,KS}.

In this paper we consider the partition function $I_b$ of $\N=4$
supersymmetric Yang-Mills matrix model obtained by the dimensional reduction to
$0$ dimensions from $D=4$ $\mathcal{N}=1$ supersymmetric Yang-Mills gauge theory with arbitrary
simple gauge group $G$
\begin{equation}
\label{Ib} I_b(\g) = \f 1 {\vol (G/Z)} \int d\l \,dA\, d D e^{-S}
\end{equation}
where the action is
\begin{equation}
S_{YM}=-\tr \lb\f 1 4 [A_{\mu}, A_{\nu}]^2 + \bar{\l}
\bar{\sigma}^{\mu} [A_{\mu}, \l] - 2D^2 \rb
\end{equation}
By $A_{\mu}$, $\lambda$ and $D$ we denote the dimensional
reduction to 0 dimension of 4-dimensional vector, 2 component
complex Weyl spinor and scalar respectively. All fields are in the
adjoint represantation of $G$, i.e. their components belong to the
Lie algebra $\g$ of the group $G$. The functional integral that
defines the dynamics of the theory is reduced to an ordinary
finite dimensional integral. This means a tremendous
simplification for a computation, nevertheless the reduced theory
remains nontrivial due to the quartic commutator potential. In
this work we will focus on computing the partition function with
zero sources (\ref{Ib}). The measure of integration is obtained
from a Killing form on $\g$. It is unique up to an overall
scaling, but the factor $\f 1 {\vol(G/Z)}$ (where $Z$ is the
center of $G$) in front of the integral makes the whole expression
invariant to rescaling and a topological structure of $G$. So, the
only argument of $I_b$ is the Lie algebra $\g$.

It should be mentioned, that in the euclidian signature
(where we are working) the integral (\ref{Ib}) is convergent.
Indeed, after integrating over $\lambda$ and $D$ one obtain a homogeneous
function of $A_{\mu}^a$, and then in the spherical coordinate system
an integration over the radius $|A|$ can be easily performed. The
obtained function of angular coordinates will be
singular at some varieties corresponding to the directions where
the potential $[A^{\mu},A^{\nu}]^2$ vanishes. Simple counting
of powers of the divergence with respect to the corank of these
 varieties shows that, nevertheless, the integral $I_b$ is convergent
\cite{Austing:2001ib}. In the simplest case of $SU(2)$ gauge group
the integral $I_b$ can be directly computed without any deformation
of the integrand \cite{Yi:1997eg,Sm2,Sethi} (and the result is
equal to $1/4$). In the case of higher rank simple groups the
integral is too sophisticated to be computed directly.
In series of works \cite{Staudacher:2000gx} a numerical Monte-Carlo
method was applied to approximately calculate $I_b$ and the result
has been obtained for all simple Lie groups with $\rk \leq 3$.

In the paper by G.~W.~Moore, N.~Nekrasov and S.~Shatashvili (MNS) \cite{MNS} the authors
managed to reduce the integral
$I_b(\su(N))$ to a contour integral of a rational function using
a certain deformation of the integrand and a localization principle that we
will briefly review in the section 2. Their method is easily
generalized for arbitrary simple Lie group. However, as it will be
seen below, the case of $SU(N)$ group is  distinguished by
existence of "the determinant formula"
\begin{footnote}
{A bosonic-fermionic correspondence for the correlator of $2N$ primary
fermionic fields $\psi(x_i)$ and $\psi^{+}(y_i)$ in conformal
field theory.}
\end{footnote}
\begin{equation}
\label{detf}
 \f {\prod_{1 \leq i<j \leq N}(x_i-x_j)(y_i-y_j)}{\prod_{1\leq i,j \leq N}(x_i-y_j)}
 = \det\nolimits_{ij} \lb \f {1} {x_i - y_j} \rb
\end{equation}
which drastically simplifies a computation of the contour integral
and eventually allows to obtain the result
\begin{equation}
I_b^{MNS}(\su(N))=1/{N^2}
\end{equation}
In spite of the fact that the relevant generalization of
(\ref{detf}) to other Lie groups is not known to us, in this work
we will try to compute the contour integrals $I_b^{MNS}(\g)$ in the
case of arbitrary simple Lie algebra $\g$.

Another approach to the problem of computation $I_b$ exists and it
is connected with  a notion of the Witten index \cite{Wit82} for a
supersymmetric quantum mechanics obtained by the dimensional reduction to one
dimension of four dimensional $D=4$ $\N=1$ super Yang-Mills
theory. The Witten index for quantum mechanics
 is defined as follows \cite{Wit82}
\begin{equation}
\label{defind} \ind = \lim_{\beta \to \infty} \tr (-1)^F e^{-\beta
H} = n^0_b -n^0_f
\end{equation}
and it counts the difference between the numbers of normalizable
bosonic and fermionic ground states. In supersymmetric quantum
mechanics with discrete spectrum $\ind$ does not depend on
$\beta$, since bosonic and fermionic states come in pairs for every
energy level $E>0$ and their contribution to $\ind$ cancel.
Therefore, instead of the limit $\beta \to \infty$ one can take
$\beta \to 0$ and compute more simple quantity
\begin{equation}
\label{ind0} \lim_{\beta \to 0} \tr (-1)^F e^{-\beta H}
\end{equation}
In terms of path integral
\begin{footnote}
{with time $t \in [0..\beta]$  and periodic boundary conditions
with respect to it}
\end{footnote}
the limit $\beta \to 0$ means reduction of $D=1$ theory to $D=0$
theory, and the quantity (\ref{ind0}) is given exactly by the
finite dimensional integral $I_b$ that we are interested in.

However, in our case this trick does not work, since the spectrum of the SYM quantum mechanic that
we consider is not discrete and $\tr (-1)^F e^{-\beta H}$ does depend on $\beta$. In
the case of continuous spectrum the discrete sum over energy
levels in (\ref{defind})
\begin{equation}
\sum_{k}{(-1)^{F_k}e^{-\beta F_k}}
\end{equation}
is replaced by an integral containing difference between bosonic
and fermionic densities,
\begin{equation}
\int_{0}^{\infty} dE {e^{-\beta E}(n_b(E) - n_f(E))}
\end{equation}
which, generally speaking, is not zero. A convenient way to
represent $\ind$ is the following \bqa \label{setup}
\ind = \lim_{\beta \to \infty} \tr (-1)^F e^{-\beta H} = I_b - I_d \\
I_b = \lim_{\beta \to 0} \tr (-1)^F e^{-\beta H} \\
I_d = -\int_{0}^{\infty} d \beta \f d {d \beta} \tr (-1)^F
e^{-\beta H} \eqa
where $I_b$ is known in literature as the bulk
contribution ("the principal term") and $I_d$ as the boundary
contribution ("the deficit term"). The value of $I_d$ has been
rigorously computed only in the case of $SU(2)$ group in
\cite{Sethi}.
\begin{equation}
I_d(\su(2))=1/4
\end{equation}
 In \cite{Yi:1997eg}
it was argued that the only contribution to $I_d$ comes from a
region of large $A^{i}$ where the initial
non-abelian $SU(2)$ theory can be approximated by a free effective
abelian U(1) theory of particles propagating along "the flat valleys"
of the commutator potential $[A^{i},A^{j}]^2$
\begin{equation}
I_d = I_d^{eff}
\end{equation}
In other words, after gauge fixing all $X^{i}$ runs
mainly along Cartan subalgebra of the gauge group, while
fluctuations in all transverse directions are suppressed at large
$X^i$ by steep walls of the potential $[X^{i},X^{j}]^2$.
Then, the effective  theory is a theory of free particles propagating on the moduli space $R^{d-1}/Z_2$.
In this  free theory $\ind^{eff}=0$, while principal contribution $I_{b}^{eff}$ can
 be easily calculated. In this way the result $I_d=I_d^{eff}=I_b^{eff}=1/4$
 was obtained in \cite{Yi:1997eg} and it coincided with the rigorous result of \cite{Sethi}.
 This fact motivated the authors of \cite{GrGut} to suggest that this prescription generalizes
 for $SU(N)$ and they obtained
 \begin{equation}
 I_d^{eff}(SU(N)) = I_b^{eff}(SU(N)) = \langle \Psi (-1)^F e^{-\beta H} \mathcal {P}\Psi \rangle = 1/N^2
 \end{equation}
In this equation $\mathcal{P}$ is the projector on the gauge
invariant states. The gauge group for this effective free theory
is the Weyl group of $G$ and thus $\mathcal{P}$ is equal to the following sum over it
\begin{equation}
\mathcal{P}=\f 1 {\#W} \sum_{w \in W} M_{w}
\end{equation}
where $M_{w}$ represent an action of a Weyl group element $w \in W$ on fields.
The obtained result $I_d(\su(N))=1/N^2$ agrees with the result in
\cite{MNS} $I_b^{MNS}(SU(N))=1/N^2$, provided $\ind=0$. In the
paper by V. Kac and A. Smilga \cite{KS} the method of \cite{GrGut}
was directly generalized for arbitrary simple Lie group and the
result was
\begin{equation}
\label{KS1} I_d^{KS}(G) = \f 1 {\# W(G)} \sum_{w \in W(G)}' \f 1
{\det (1-w)}
\end{equation}
For $SU(N)$ this expression simplifies to $1/N^2$ and agrees with
the results of MNS \cite{MNS} and with  direct numerical
computation of $I_b$ in \cite{Staudacher:2000gx}. However, in
\cite{Staudacher:2000gx} the integral $I_b$ has been calculated
for all simple Lie groups of $\rk \leq 3$ both by numerical
(Monte-Carlo) and  MNS \cite{MNS} methods. While numerical and
MNS contour integral results coincided with each other, they did
not agree with KS
 formula (\ref{KS1}). This strongly indicates that the free effective
 hamiltonian method exploited in \cite{KS} fails to calculate $I_d$.

\section{Localization}

We are going to exploit the method used by MNS \cite{MNS} to
compute the integral (\ref{Ib}) for arbitrary simple Lie group.

Let us briefly review how supersymmetry allows us to greatly
simplify an exact analytical calculation of the partition
function. An integral like $\int D \phi e^{-S[\phi]}$ can be
approximately computed using the quasiclassic approximation as the sum
$(\det_{ij} \f {\d^2 S} {\d \phi_i \d \phi_j})^{-1/2}$ over
the critical points $\f {\d S} {\d \phi} =0$ of the action $S$.
\begin{footnote}{
Assuming the measure $\int D \phi e^{-\phi^2}=1$}
\end{footnote} It often happens that some symmetry makes all
corrections to this approximation vanish. In other words, it means
that the integral $\int D \phi e^{-S[\phi]}$ is localized  on
the critical points of the action.
\begin{footnote}
{A toy example of such a localization is the following integral:
 $\int_{S^2} d \Omega e^{-\beta \cos \theta} =  \f {2 \pi} {\beta}
 (e^{\beta}-e^{-\beta})$}
\end{footnote}

To see how it happens  one can deform the integrand by a full
derivative in such a way that makes the integral nearly gaussian.
Suppose, that $S$ is invariant under the
action of some nilpotent derivative
\begin{footnote}{
$Q(AB)=(QA)B \pm A(QB)$}
\end{footnote}
 operator $Q$
\begin{equation}
\label{Qcond} QS=0, \qquad Q^2=0
\end{equation}
Integration by parts shows how the partition function
changes if the action is deformed by a $Q$-exact term:
\bqa
\label{locex}
Z(g)= \int d \phi \, e^{-tS(\phi)+gQW(\phi)} \\
\d_g Z = \int d \phi \, e^{-tS(\phi)+gQW(\phi)} QW(\phi)= \\=
\label{qq} \int d \phi \, Q ( W(\phi)e^{-tS(\phi)+gQW(\phi)})-\int
d \phi \,W(\phi)Q e^{-tS(\phi)+gQW(\phi)}
\eqa
The second term of
(\ref{qq}) vanishes due to (\ref{Qcond}). The first one can be
rewritten as a boundary contribution
\begin{equation}
\label{bc} \int d\phi \f {\d} {\d \phi} (
W(\phi)e^{-tS(\phi)+gQW(\phi)} Q\phi )
\end{equation}

This terms also vanishes if a domain of integration is a compact
space without a boundary. In the case of a non-compact space one
can neglect a boundary contribution if the integrand of
(\ref{bc}) decreases exponentially fast when $\phi$ tends to
the infinity. If an action itself is $Q$-exact, then we can send the
coefficient $t$ in front of the action in the exponent in \ref{locex} to the
infinity, since that does not affect the value of the integral.
 In such a way it turns into the gaussian one and therefore is equal to
the sum over critical points, as claimed above.

Let us apply this method to compute $I_b$ (\ref{Ib})
\cite{MNS}.

The SYM action
\begin{equation}
S_{YM}=-\tr \lb\f 1 4 [A_{\mu}, A_{\nu}]^2 + \bar{\l}
\bar{\sigma}^{\mu} [A_{\mu}, \l] - 2D^2 \rb
\end{equation}
has  $\mathcal{N}=4$ $D=0$ supersymmetry transformations which
 can be obtained by the dimensional reduction
 from $\N=1$ supersymmetry transformation in $D=4$ dimensions:
\bqa
\vd_{\ez} A^{\mu} &=& -i \bar{\l} \bar{\s}^{\mu} \ez + i \bar{\ez} \bar{\s}^{\mu} \l \\
\vd_{\ez} \l &=& i \s ^{\mu,\nu} \ez [A^{\mu}, A^{\nu}] - 2 \ez D \\
\vd_{\ez} D &=& \f 1 2 [A^{\mu},\bar{\l}] \bar{\s}^{\mu} \ez + \f
1 2 \bar {\ez} \bar{\s}^{\mu} [A^{\mu},\l] \eqa where $\s^{\mu\nu}
= \f 1 4 (\s^{\mu} \bar{\s}^{\nu} - \s^{\nu} \bar{\s}^{\mu})$.

To rewrite the action and the supersymmetry transformation in a more
convenient way we make change of variables as following:
\begin{equation}
\bal \l_1 &= \eta + i \psi \qquad
&\l_2 = \chi_1 - i \chi_2 \\
u &= \f 1 2 (A^3 + i A^4) \qquad
&\bar{u} = \f 1 2 (A^3 - i A^4) \\
X^{1} &= A^{1} \qquad
&X^{2} = A^{2} \\
D &= H + \f 1 2 [A^{1},A^{2}] \eal
\end{equation}

Then we choose one of four supersymmetry generators $\vd$ that
acts as following: \bqa
&\vd \psi = H \qquad & \vd H= [u,\psi] \\
&\vd \ub = \eta \qquad & \vd \eta = [u,\ub] \\
&\vd X^i = \chi^i  \qquad & \vd \chi^i = [u,X^i] \\
&\vd u =0 \label{s1} \eqa

It can be easily checked that  the action can be rewritten as a
$\vd$-exact expression
\begin{equation}
\label{S0}
 S = \vd \lb [u,\ub] \eta + [X^i,\ub] \chi^i +[X^i,X^j] \psi \ev_{ij}+ H \psi \rb
\end{equation}

The supersymmetry $\vd$ squares to a gauge transformation $\vd^2 =
[u,\cdot]$ generated by $u$ and thus $\vd^2=0$ on gauge invariant
quantities.

Therefore, the above described philosophy is suitable in our case:
the integrand is localized on classical equations of motion (or
a moduli space of the low energy effective theory). However, an
additional comprehension arises because this moduli space is not
compact and an integration over it is ill defined. In \cite{MNS}
it was suggested to slightly deform the supersymmetry generator
$\vd \to \vd_{\e}$ in  such a way that $\vd_{\e}$ squares both to
the gauge transformation $[u,\cdot]$ and $Spin(2)$ Lorentz
rotation $T_{\e}(Spin(2)_{X_1,X_2})$ in the plane $(X_1,X_2)$
\bqa
&\vd_{\e} \psi = H \qquad & \vd_{\e} H= [u,\psi] \\
&\vd_{\e} \ub = \eta \qquad & \vd_{\e}\eta = [u,\ub] \\
&\vd_{\e}X^i = \chi^i  \qquad & \vd_{\e}\chi^i = [u,X^i] + \e
\ev_{ij} X^j \label{s2}
\eqa
The deformation method still works if
we deal with Lorentz and gauge invariant expressions. After
the deformation $\vd \to \vd_{\e}$  the additional term  $\e
[\bar{u},X^{i}] \ev_{ij} X^j$  is added to the action. By scaling
of variables of integration it can be shown that the integral does
not depend on $\e$. If we also require the limit $\e \to 0$
not to be singular, then the value of the deformed integral will not
change.

 So, we continuously deform the original $\vd$-exact  action (\ref{S0}) to another
 one, more suitable for computation:
\begin{equation}
 S_{\e}^{def} = \f 1 g \vd_{\e} (X_i \chi_j \ev_{ij} + \ub \psi) =
  \f 1 g \left[
 (\chi_i \chi_j  + X_i[u,X_j])^{\phantom{\frac12}}\ev_{ij} - \e X_i X_i+
 ( \eta \psi + \ub H ) \right]
\end{equation}

Then the trivial gaussian integration on $X^i,\chi^i, \ub, \eta, H,
\psi$ gives
\begin{equation}
\bal
 I_b = \f 1 {\vol (G/Z)} \int du \, dX^i d\chi^i \, dH d\psi \, d\ub d\eta \
 e^{\f 1 g \left[
 (\chi_i \chi_j  + X_i[u,X_j])\ev_{ij} - \e X_i X_i +
 ( \eta \psi + \ub H ) \right] } =\\
 =\f 1 {\vol (G/Z)} \int du \, \f 1 {\det (ad(u)+\e)}\\
\eal
\end{equation}

Now we see, that the deformation $\vd \to \vd_{\e}$ resolves
the singularity $\det (ad(u))=0$
\begin{footnote}
{$\det (ad(u))=0$ since the adjoint action of $u$ vanishes on
elements from Cartan subalgebra $\mathfrak{h}$}
\end{footnote}
 by introducing into the action the mass term for $X_1,X_2$.
Then, using the gauge invariance one can reduce the integration
over the whole Lie algebra $\mathfrak{g}$ to the integration over
its r-dimensional Cartan subalgebra $\mathfrak{h}$
\begin{equation}
\bal I_b^{MNS}(G) = \f 1 {\vol (G/Z)} \int_{\mathfrak{g}} du \, \f
1 {\det (ad(u)+ \e)} =
 \f {1} {\#W} \int_{\mathfrak{h}} \f {d^r u} {\vol(T/Z)}\, \f {\det'(ad(u))}
  {\det (ad(u) + \e)}
\eal
\end{equation}

By $\#W$ we denote the order of the corresponding Weyl group and by $\#Z$ the
order of the center of $G$. The factor ${\det'(ad(u))}$
\begin{footnote}
{by $\det'(ad(u))$ we denote the determinant of $ad(u)$ acting on
$\mathfrak{g} \setminus \mathfrak{h}$}
\end{footnote}
 in the numerator appears from the volume of the orbit of $u \in \mathfrak{h}$
 obtained by a $G$-adjoint action.

Then we can explicitly rewrite the integral
 only in terms of roots $\{\alpha\}$ of the algebra~$\g$
\begin{equation}
\label{contur}
\bal I_b^{MNS}(\g) = \f {\det \| \a^s_{ij} \|} {\#W} \int_{\cc} du_1 \int_{\cc} du_2 \dots \int_{\cc} du_r \, \f 1 {(2 \pi i
\e)^r} \prod _{\a} \f {\a u } {\a u +\e} \eal
\end{equation}
(where $\{\alpha^s\}$ is the set of simple roots).

The performed deformation is valid if the domain of integration is
compactified and  the integrand is not singular on it. The compact
domain of integration implies that the point $u=\infty$ should be included to it,
and the reiterative integrals in (\ref{contur}) should be taken along the closed
contour $\cc$. The integrand in (\ref{contur}) has poles at the points
$\a u  + \e = 0$ and $u=\infty$, and contour of integration
should not pass throw them. Therefore, we shift the initial
contours of integration for  $u^i \in \mathbb{\overline{C}}$ from the point $u^i=\infty$.
In terms of a complex plane $\mathbb{C}$ this means that the deformed contours of
integrations for every $u^i$ are now consist of the real axis $\mathbb{R} \subset \mathbb{C}$ and
are closed by the
infinite upper or lower arc.
\begin{footnote}
{ It really does not matter whether we close contours by the upper
or the lower arc, but it is necessarily to close contours in the same
manner for all $u^i$}
\end{footnote}

Despite the performed reduction of the initial integral $I_b$ to
the contour integral $I_b^{MNS}$ might seem to be not rigorous
enough, we do emphasize that it was strongly supported by the direct
numerical (Monte-Carlo) evaluation of the integral $I_b$ for all
simple groups of $\rk \leq 3$ in \cite{Staudacher:2000gx}.

\section{Evaluating of the contour integral $I_b^{MNS}(\g)$}

The contour integral (\ref{contur}) of a rational function can be
computed by residues and it was done in the original work
\cite{MNS} for $SU(N)$.

The crucial step in that computation was to represent a large
product over the set of roots of $A_{N-1}$ as a sum over
permutations (it is the $SU(N)$ Weyl group) of rather simple terms
with a help of "the determinant formula" that was mentioned in the
Introduction.
\begin{equation}
 \f 1 {\e^r} \prod_{i\neq j} \f {u_i - u_j} {u_i - u_j + \e } =
 \sum _{\sigma \in \Ss_N} \prod_{i=1..N} \f {1} {u_{i} - u_{\sigma(i)}+\e}
\end{equation}
After evaluation of contour integral by residues it can be seen,
that only terms corresponding to the longest cycles (there are
$(N-1)!$ such cycles) of permutations $\Ss_{N}$ remain. Each such term
 contributes to the sum $1/N^2$. Thus, the MNS result is
\begin{equation}
I_b^{MNS}(SU(N)) = \f N {N!} \f  {(N-1)!} {N^2} = \f 1 {N^2}
\end{equation}
The additional factor $N$ in the numerator is the order of the
center $Z_N$ of $SU(N)$.

Our task is to evaluate the integral $I_b^{MNS}(\g)$ for arbitrary
simple Lie algebra. The direct extension of the MNS method for
arbitrary simple Lie algebra would be to find an analogue of
"the determinant formula" that represent product over all roots of
an algebra $\g$ in equation (\ref{contur}) as a sum over its Weyl group
of simpler terms,
\begin{footnote}
{These terms may be products over some subsets of the set of roots}
\end{footnote}
suitable for a computation of the contour integral.
Unfortunately, we have not managed to find  an analogue that would be relevant in our case.
For the exceptional groups the integral $I_b^{MNS}$ can be explicitly
evaluated, and it was done in \cite{Staudacher:2000gx} for $G_2$. We
computed also $I_b^{MNS}$ for $F_4$ and $E_6$ and so only $E_7$
and $E_8$ values are still unknown now.

As regards the other infinite classical series $B_N,C_N,D_N$ we explicitly
computed $I_b^{MNS}(\g)$ for $N \leq 11$ and conjectured general
formulas for every $N \in \mathbb{N}$. For all groups, except those
that are isomorphic to the unitary one, our results do not coicide
with the $KS$ expression (\ref{KS1}). However,  we explicitly
evaluated the KS sums (\ref{KS}) over Weyl groups for the classical series
$B_N,C_N,D_N$ and can provide a suggestion on how KS formula could
be modified to give results agreeing with $I_b^{MNS}$.

\section{Explicit results}
To our best knowledge of the literature, it seems that the reliable results for the matrix
integral $I_b$ have been obtained before for the following groups
\begin{itemize}
\item $SU(N),
\begin{footnote}{$A_N$ - series} \end{footnote}
\quad \forall N$ \cite{MNS} \item $SO(2N), SO(2N+1), Sp(2N),
\begin{footnote}{$D_N,B_N,C_N$ - series} \end{footnote}
\quad N \leq 3$ \cite{Staudacher:2000gx} \item $G_2$
\cite{Staudacher:2000gx}
\end{itemize}
So, the entire infinite series $B_N,C_N,D_N$ and the remaining exceptional groups ($F_4, E_6,
E_7, E_8$) have to be explored to complete the story.

In addition to the previous results, in this work we explicitly
computed the contour integral (\ref{contur}) for the following
groups
\begin{footnote}{
A special symbolic manipulation program was written to sum up a
huge number of residues (its number grows faster then $N!$ and so
we are limited by rather small $N$)}
\end{footnote}
:

\begin{itemize}
\item $Sp(2N), SO(2N), SO(2N+1), \quad N \leq 11$ \item $F_4$
\item $E_6$
\end{itemize}

Here is the table of our results for the series $B_N, C_N, D_N$:

\bigskip
\begin{tabular}{|l|l|l|l|}
\hline
N & $Sp(2N) \quad [C_N]$  & $SO(2N+1) \quad [B_N]$ & $SO(2N) \quad [D_N]$\\
\hline
1 & 1/4 & 1/4 & 0 \\
2 & 9/64 & 9/64 & 1/16 \\
3 & 51/512 & 25/256 & 1/16 \\
4 & 1275/16384 & 613/8192 & 117/2048 \\
5 & 8415/131072 & 1989/32768 & 53/1024 \\
6 & 115005/2097152 & 26791/524288 & 6175/131072 \\
7 & 805035/16777216 & 92599/2097152 & 5661/131072 \\
8 & 45886995/1073741824 & 5220675/134217728 & 1338019/33554432 \\
9 & 331406075/8589934592 &  18671491/536870912 & 310819/8388608 \\
10 & 838528695/137438953472 & 270276175/8589934592 & 74352375/2147483648 \\
11 & 35629165845/1099511627776 & 987486975/34359738368 & 69819475/2147483648 \\
\hline
\end{tabular}

\bigskip

and for the exceptional groups
\begin{footnote}{
The first fraction of these numbers is $\f 1 {\#W}$}
\end{footnote}

\bigskip
\begin{tabular}{|l|l|}
\hline
$G_2$ & $ \f {1}{12} \f {151} {72} $\\
$F_4$ & $ \f 1 {1152} \f {493013} {3456}  $ \\
$E_6$ & $ \f 1 {51840} \f {286340} {81}    $ \\
\hline
\end{tabular}

\bigskip

The first lines of these tables $N \leq 3$ are in a perfect agreement with
the previous results of \cite{Staudacher:2000gx} where similar integrals for
all simple algebras with $\rk \leq 3$ have been computed.


Our conjectured generic formulas for the classical infinite series
$B_N,C_N,D_N$ can be found in the equations
(\ref{so2n1imp}),(\ref{ResSp2N}),(\ref{so2nimp}) below.

\section{The deficit terms $I_d$ and the conjecture for a general formula for $I_b^{MNS}$
 in the case of $SO(2N+1),Sp(2N+1),SO(2N)$ groups}

In \cite{KS} the following formula for $I_d(\g)$ was suggested.
\begin{equation}
\label{KS} I_d^{KS}(\g) = \f 1 {\# W} \sum'_{w \in W} \f 1 {\det
(1-w)}
\end{equation}
As it was stated above this expression does not agree with the
result obtained by the MNS deformation technique
\begin{footnote}{
but that results are strongly supported by numerical Monte-Carlo evaluation in \cite{Staudacher:2000gx}.}
\end{footnote}
for all simple groups except the unitary one (or isomorphic
to it).
\begin{footnote}
{The fractional parts of $I_b$ and $I_d$ should coincide
regardless of the value of $\ind$ (since it is integer), but they do not. In the following, we will
conjecture that $\ind=0$ since in all known cases $I_b < 1$, $I_d > 0$ and $0 \leq \ind=I_b - I_d$ }
\end{footnote}

In the case of the unitary group $SU(N)$ the expression (\ref{KS}) can
be easily computed. Indeed, in the sum over permutations $\Ss_N$
only the longest cycles contribute $\det(1-w)^{-1}=1/N$, while the
others have $\det(1-w)=0$ and thus are thrown away. There are $(N-1)!$
longest cycles. So, the result is
\begin{equation}
I_d^{KS}(\su(N)) = \f 1 {N!} \f  {(N-1)!} {N} = \f 1 {N^2}
\end{equation}
and it is correct.

Let us consider now the cases of the classical series $B_N,C_N,D_N$, where
the KS formula (\ref{KS}) does not agree with the MNS contour integral
(\ref{contur}) in known examples.

We have managed to explicitly compute the KS formula (\ref{KS}) as
a function of rank $N$ for all classical series
($SO(2N+1)$,$Sp(2N)$, $SO(2N)$).

Firstly, let us consider $C_N$-case ($Sp(2N)$).
 The $C_N$ Weyl group $W$ consists of permutations and changing signs of $u_i$.
\begin{equation}
 W = (Z_2)^N \ltimes \mathcal{S}_N
\end{equation}

We can classify all elements of  $W$ in the following
way.  Every permutation can be broken into the direct product
 of $l$ cycles $c_{k_i}$ of lengths $k_1,k_2\dots k_l$.
 \begin{equation}
 \det(1-w)=\prod_{i=1..l} \det(1-c_{k_i})
\end{equation}

Then, for every cycle $c$ we note the following identity
\begin{equation}
\det(1-c) = 1 + \det(-c)
\end{equation}
(Proof. Consider the determinant of the corresponding matrix $1-c$
as a tautological sum over all permutations of products
$\prod_{i=1..k} (1-c)_{i,\sigma(i)}$. In this sum only two terms
corresponding to the product of $1$ on the main diagonal and
elements of matrix $c$ contribute).

The determinant of cycle can be easily computed:
\begin{equation}
\det(-c_{k_i}) = (-1)^{k_i}(-1)^{k_i+1}(-1)^{\#-} =  (-1)^{1+\#-}
\end{equation}

Thus, $\det(1-c)$ is equal to $2$ or to $0$ in equal number of
cases.
\begin{footnote}{
It is non zero when the number of minus signs is odd. It means
that this element of the Weyl group considered as a permutation of
$2k_i$ elements $\{\pm u_i\}$ is a cycle of length $2k_i$ and cannot be broken
into two disjoint cycles of length $k_i$ as it would when $\#-$ is
even. }
\end{footnote}

A number of ways of choosing long cycles $k_1,\dots k_l$ is $ \f
{N!} {k_1!\dots k_l!}$. Each cycle $k_i$ can be realized in
$2^{k_i}(k_i-1)!$ ways and only a half of them contribute to the
corresponding term an additional factor $1/2$ each.

So, we obtain
\begin{equation}
\bal \label{explsp2N}
\sum'_{w \in W_{Sp(2N)}} \f 1 {\det (1-w)} =\\
= \sum_{l=1}^{N} \f 1 {l!} \sum_{k_1+k_2+\dots+k_l=N} \f
{N!}{k_1!k_2!\dots k_l!} (k_1-1)!(k_2-1)!\dots(k_l-1)!
\f {2^{k_1+k_2+\dots+k_l}} {2^l 2^l} =\\
=2^{N} \sum_{l=1}^{N} \f {2^{-2l}} {l!} \sum_{k_1+k_2+\dots+k_l=N}
\f {N!}{k_1 k_2 \dots k_l} \eal
\end{equation}

Then, using the identity
\begin{equation}
 \f 1 {x_1 x_2 \dots x_N} = \sum_{\sigma \in S_N} \f 1 {x_{\sigma(1)}
({x_{\sigma(1)}}+{x_{\sigma(2)}}) \dots
({x_{\sigma(1)}}+\dots+{x_{\sigma(N)}})}
\end{equation}
and making change of variables $k_i \to \sum_{j=1}^{i} k_j$, we
obtain the following expression for the number of permutations
consisting of $l$ disjoint cycles (it is an absolute value of the
Stirling number $S(N,l)$ of the first kind)
\begin{equation}
\bal F^l_N \equiv \f 1 {l!} \sum_{k_1+k_2+\dots+k_l=N} \f {N!}{k_1
k_2 \dots k_l} =
\sum_{k_1+k_2+\dots+k_l=N} \f {N!} {k_1(k_1+k_2)\dots(k_1+\dots+k_l)}=\\
=\sum_{1\leq k_1 <k_2<\dots<k_l=N} \f {N!} {k_1 k_2 \dots k_l}=\\
= \sum_{1\leq k_1 <k_2<\dots<k_{N-l}\leq N-1} k_1 k_2 \dots
k_{N-l} \eal
\end{equation}

Finally, we obtain
\begin{equation}
\bal \label{weylexp} \sum_{w \in W_{Sp(2N)}}' \f 1 {\det (1-w)} =
\sum_{l=1}^{N} 2^{N-2l} F^l_N=2^N \sum_{l=1}^{N} \f {1} {2^{2l}}
\sum_{1\leq k_1 <k_2<\dots<k_{N-l}\leq N-1}
 k_1 k_2 \dots k_{N-l}
=\\= 2^N \prod_{k=0}^{N-1} (k+\f 1 4) = 2^{-N} \prod_{k=0}^{N-1}
(4k+1) \eal
\end{equation}
So, the $Sp(2N)$ KS formula (\ref{KS}) is reduced to the following function of $N$
\begin{equation}
\label{KSSp2N} I_d^{KS}(\sp(2N)) = \f 1 {2^{2N} N!}
\prod_{k=0}^{N-1} (4k+1)
\end{equation}

This formula provides the correct value $1/4$ for $N=1$ (in this case
$Sp(2)$ is isomorphic to the unitary group $SU(2)$),
 but does not coincide with the contour integral $I_b^{MNS}$ for all other $N>1$.
 For  $N=2$ we have

\begin{equation}
\#W(C_2) \, I_d^{KS}(\sp(4)) = 1+1/4=5/4,
\end{equation}

while

\begin{equation}
\#W(C_2) \, I_b^{MNS}(\sp(4)) = 9/8
\end{equation}

However, we note, that the results for the contour integral
 $I_b^{MNS}(\sp(2N))$ obtained by the explicit calculation
for $N \leq 11$ are perfectly described by the following
function
\begin{equation}
\label{ResSp2N} I_b^{MNS}(\sp(2N)) = \f 1 { 2^{3N-1} N!}
\prod_{k=0}^{N-1} (8k+1)
\end{equation}
which is different, but still rather similar to (\ref{KSSp2N}).

If one converts this expression into a form similar to the
(\ref{weylexp}), it can be seen that to make KS formula
(\ref{KS}) agree with $I_b^{MNS}$
 in the $Sp(2N)$ case it is sufficient to multiply terms in the sum corresponding to
 $l$ disjoint permutation cycles by an additional factor $1/{2^{l-1}}$. We can interpret
this correction in the sense that every additional disjoint
permutation cycle contributes another factor $1/2$ into the
products in the sum (\ref{weylexp}) $ \sum_{l=1}^{N} 2^{N-2l}
F^l_N \longrightarrow  \sum_{l=1}^{N} 2^{N-2l} 2^{1-l} F^l_N$.

It is also a curios fact that in both cases of $SU(N)$ and $Sp(2N)$
groups
\begin{footnote}{But not for every
 simple Lie group, of course!}\end{footnote}
 the following expression for $I_d$ agrees with $I_b^{MNS}$
\begin{equation}
\label{KSmod} I_d (G) = \f {\# Z} {\# W} \sum'_{w \in W} \f 1
{\det (1-w)^2}, \qquad G \in \{SU(N),Sp(2N)\}
\end{equation}

 We have also managed to explicitly compute the KS formula (\ref{KS}) for
 the $D_N$ Weyl group. The $D_N$ Weyl group can be realized in the same
manner as the $B_N$ Weyl group with an additional requirement that a
number of changed signs $u_i \to -u_i$ should be even. In order to calculate
(\ref{KS}) we now need to sum up only over permutations that consist
of cycles with an even number of minus signs. That means that the terms
with an odd number of cycles in the permutation should be projected out. So, we obtain

\begin{equation}
\label{KSmod11} \sum'_{w \in W_{SO(2N)}} \f 1 {\det (1-w)} =
\sum_{l=1}^{N} 2^{N-2l} F^l_N \f {(1+(-1)^l)} {2} = 2^{-N-1} \lb
\prod_{k=0}^{N-1} (4k+1) + \prod_{k=0}^{N-1} (4k-1)\rb
\end{equation}
and finally
\begin{equation}
I_d^{KS}(\so(2N)) = \f {2} {2^{N-1}N!}2^{-N-1} \lb
\prod_{k=0}^{N-1} (4k+1) + \prod_{k=0}^{N-1} (4k-1)\rb
\end{equation}
Again, KS formula $I_d^{KS}$ (\ref{KS}) does not agree with the MNS
contour integral $I_b^{MNS}$. But we observe that the result for
the MNS contour integral (\ref{contur}) in the explicitly computed  $SO(2N+1)$ and
$SO(2N)$ ($N \leq 11$) cases can be obtained by multiplying the terms in the sum
(\ref{weylexp}) and in the middle part of (\ref{KSmod11})
respectively  by the same factors from the sequence
$b_l=\{1,1,-2,-2,76,76,\dots \}$.

\bqa \label{so2nimp}
I_b^{MNS}(\so(2N+1)) = \f {1} {\# W}  \sum_{l=1}^{N} 2^{N-2l} 2^{1-l} F^l_N  \,\, b_l \\
\label{so2n1imp} I_b^{MNS}(\so(2N)) =   \f {2} {\# W}
\sum_{{\bf even} \, l=2}^{N} 2^{N-2l} 2^{1-l} F^l_N  \,\, b_l \eqa

Obviously,  given arbitrary sequence of numbers $X_N$
\begin{footnote}
{In the following we will substitute $X_N$ by $I_b^{MNS}(\so(2N+1))$ or $I_b^{MNS}(\so(2N))$ }
\end{footnote}
and sequences $\{f^l_N \neq 0 , \enskip l=\overline{1 \dots N}\}$
\begin{footnote}
{by $f^l_N$ we mean $\f {\# Z} {\# W} 2^{N-2l} 2^{1-l} F^l_N $ }
\end{footnote}
we can always linearly expand $X_N$ over the set $f^l_N$
\begin{equation}
 \sum_{l=1}^N f^l_N b_l  = X_N
\end{equation}
The coefficients $b_l$ can be recurrently found: \bqa
b_1=X_1/f_1 \\
\vdots \\
b_N=(X_N-\sum_{l=1}^{N-1} f_l b_l)/f_N \eqa

In this way we can unambiguously get the two sequences $\{b(SO(2N))\}$
and $\{b((SO(2N+1))\}$. Then, we note that up to $N\leq 11$
\begin{enumerate}
\item{ These two sequences $$\{b_{2k}(SO(2N))\}$$
$$\{b_{2k}(SO(2N+1))\}$$  have coincided with each other in spite of the
fact that the algebras, root systems and the contour integral expressions \ref{contur} are
different!
\begin{equation}
b_{2k}(SO(2N))\ = b_{2k}(SO(2N+1)), \quad \forall k
\end{equation}
} \item{In the $SO(2N+1)$ case when in the sum also terms with
odd $l=2k+1$ are present, the coefficients $b_{2k+1}$ are equal to the preceding ones $b_{2k}$.
\begin{equation}
b_{2k+1}=b_{2k}, \qquad \forall k
\end{equation}
} \item{ The signs of pairs ($b_{2k}=b_{2k+1}$) are interchanged.
\begin{equation}
 \sign b_{2k} = (-1)^{k+1}
\end{equation}
} Let us represent the sequence $b_{2k}$ in the following way
\begin{equation}
b_{2k} = (-1)^{k+1} 2^k \beta_k
\end{equation}
Then $\beta_k = \{1,1,19,559,29161,\dots \}$
\item The most nontrivial observation and indeed a conjecture about a value for
the MNS contour integral $\forall N \in \mathbb{N}$ in the $SO(2N)$
and $SO(2N+1)$ case is the following. The coefficients $\beta_k$
are expressed as numerators in the Taylor expansion of the generating
function $\sqrt{\cos(x)}$.
\begin{equation}
 \sqrt {\cos x } = 1  - \sum_{k=0}^{\infty} \f {\beta_k x^{2k}} {2^k (2k)!}
\end{equation}
\end{enumerate}
We conjecture that these observable properties of $b_{2k}$ and
$b_{2k+1}$ for $N\leq 11$ are extended for arbitrary $N$.

\section{Conclusion}
  In this work the MNS contour integral $I_b^{MNS}(\g)$
  has been explicitly evaluated in the case
  of classical $SO(2N+1), Sp(2N), SO(2N)$ groups for $N \leq 11$ and
also for the exceptional groups $G_2$, $F_4$, $E_6$. The
conjectures about values of $I_b^{MNS}(G)$ in the case of
classical infinite series $SO(2N+1), Sp(2N) ,SO(2N)$ have been
suggested $\forall N \in \mathbb{N}$. In addition, KS formula
(\ref{KS}) for $I_d^{KS}(\g)$ has been explicitly computed as a
function of rank $N$ in the $SO(2N+1), Sp(2N) ,SO(2N)$ cases. The
results disagreed with the values of $I_b^{MNS}$ but the computation
helped us to conjecture the value of $I_b^{MNS}(\g)$ in the case of
classical infinite series $SO(2N+1), Sp(2N), SO(2N)$.

For $Sp(2N)$ series it is simple:
\begin{equation}
I_b^{MNS}(\sp(2N)) = \f 1 { 2^{3N-1} N!} \prod_{k=0}^{N-1} (8k+1)
\end{equation}

For $SO(2N+1)$ and $SO(2N)$ series the result is build with the
help of the generating function $\sqrt{\cos(x)}$ in the following
way:

1) Define $\beta_k$ as
\begin{equation}
\sqrt {\cos x } \equiv 1  - \sum_{k=0}^{\infty} \f {\beta_k
x^{2k}} {2^k (2k)!}
\end{equation}

2) Define $b_{2k}$ and $b_{2k+1}$
\begin{equation}
b_{2k} \equiv (-1)^{k+1} 2^k \beta_k
\end{equation}

3) Define $F^l_N= S(N,l)$ as Stirling number of the first kind
\begin{equation}
F^l_N \equiv \sum_{1\leq k_1 <k_2<\dots<k_{N-l}\leq N-1} k_1 k_2
\dots k_{N-l}
\end{equation}

Then the result is

\begin{equation}
I_b^{MNS}(SO(2N+1)) = \f 1 {2^N N!} \sum_{l=1}^{N} 2^{N+1-3l}
F^{l}_{N} b_k
\end{equation}

\begin{equation}
I_b^{MNS}(SO(2N)) = \f 2 {2^{N-1} N!} \sum_{{\bf even}l=2}^{N}
2^{N+1-3l} F^{l}_{N} b_k
\end{equation}

And here are the explicit expressions for $I_d^{KS}(\g)$ for the orthogonal
and symplectic groups
\begin{equation}
I_d^{KS}(\sp(2N)) = I_d^{KS}(SO(2N+1) =  \f 1 {2^{2N} N!}
\prod_{k=0}^{N-1} (4k+1)
\end{equation}

\begin{equation}
I_d^{KS}(\so(2N)) = \f {2} {2^{N-1}N!}2^{-N-1} \lb
\prod_{k=0}^{N-1} (4k+1) + \prod_{k=0}^{N-1} (4k-1)\rb
\end{equation}

\section{Outlook or what remains to do?}

\begin{enumerate}
\item{

 To prove our conjectures about the values of $I_b^{MNS}$ for the classical series $B_N,C_N,D_N$
 for arbitrary $N$. We suggest that this part is technical
since one need to prove pure combinatorial identity. }

\item{
To understand on an algebraic level how the MNS contour integral is
connected with the KS (\ref{KS}) formula. (We observed that it is
sufficient to replace $$\det(1-w) \longrightarrow \det(1-w)^2$$
and multiply terms in the sum by some additional integer tuning
factors (like $b_k$). This procedure works for $G_2$ group too
(see Appendix). But what is hidden behind this manipulation?)}
And what exactly is wrong about the effective free hamiltonian method at
large $X_i$ in super Yang-Mills quantum mechanics?
\item{
Perhaps, it possible to find such a deformation of the initial integral
$I_b$ that will be suitable for using some generalization of
"the determinant formula" (that one also needs to find) and directly
deduce our conjectures.}
\item{
To understand in a more direct way how the generating function $\sqrt{\cos(x)}$ is connected with
$SO(2N),SO(2N+1)$ $D=0$ $\mathcal{N}=4$ super Yang-Mills matrix model.}
\item{
To accurately perform a reduction of the matrix integral $I_b$ to the
contour integral $I_b^{MNS}$ and obtain a rigorous proof of the
way that was used to deform the  integration contours}
\item{
To obtain the values of the matrix integral $I_b$ for arbitrary simple group
in the case of dimensional reduction to $0$ dimension of $D=6$ and $D=10$
super Yang-Mills. In the $SU(N)$ $D=10$ case the result was
computed in the same paper \cite{MNS}
\begin{equation}
I_b = \sum_{d \,\,| N {\rm mod} \,d = 0} \f 1 {d^2}
\end{equation}
but what about other simple groups? }
\end{enumerate}

\section{Acknowledgments}
I would like to thank R.~Anno, E.~Akhmedov, A.~Chervov,
A.~Dymarsky, A.~Gerasimov, A.~Morozov, A.~Savvateev and K.~Selivanov
for helpfull discussions, suggestions and comments. I also
acknowledge for the kind hospitality "ICTP Spring School on
superstrings and related matters" and "ITEP-ITP Kiev spring school".
This work was partly supported by the INTAS grant 00-561, the
RFBR  grant 01-02-17488a, the RFBR grant for Support of Young
Scientists 02-02-06517 and by the Russian President's grant
00-15-99296.

\appendix

\section{Explicit expressions for contour integrals}

Here we provide explicit expressions for the contour integrals
\begin{equation}
\bal I_b^{MNS}(G) = \f {1} {\vol(T/Z)\#W} \int d^r u \, \f 1 {(i
\e)^r} \prod _{\a} \f {\a u } {\a u +\e} \eal
\end{equation}
for various simple Lie algebras expressed through their root
systems. The factor $\f {1} {\vol(T/Z)}$  in front of the integral
can be rewritten as $\f {k} {(2 \pi)^r}$, where $k$ is the volume of
the cell spanned by the set of simple roots $\a^s_i$ in the weight
space:
\begin{equation}
k = \det \| \a^s_{ij} \|
\end{equation}
and
\begin{equation}
\bal I_b^{MNS}(G) = \f {k} {\#W} \int d^r u \, \f 1 {(2 \pi i
\e)^r} \prod _{\a} \f {\a u } {\a u +\e} \eal
\end{equation}

\subsection{ $SO(2N+1)$ or $B_N$}
The root system is the set
\begin{equation}
\{\a u\}=\{\pm u_i \pm u_j, \quad i<j\} \cup \{\pm u_i\}
\end{equation}
The Weyl group consists of permutations and changing signs of
$u_i$.
\begin{equation}
 W = (Z_2)^N \ltimes \mathcal{S}_N
\end{equation}
And \bqa
 &\#W = 2^N N!\\
 &k = 1
\eqa \bqa
 I_b^{MNS}(SO(2N+1)) = \f 1 {2^N N!} \f 1 {(2\pi i \e)^N} \int d^N u \prod_{i<j}
 \f {(u_i+u_j)^2(u_i-u_j)^2}  {((u_i+u_j)^2-\e^2)((u_i-u_j)^2-\e^2)}
     \prod_{i} \f {  u_i^2} { ( u_i^2-\e^2) }
 \eqa

\subsection{ $Sp(2N)$ or $C_N$}
The root system is the set
\begin{equation}
\{\a u\}=\{\pm u_i \pm u_j, \quad i<j\} \cup \{\pm 2u_i\}
\end{equation}
The Weyl group is the same as for $SO(2N+1)$. It consists of
permutations and changing signs of $u_i$.
\begin{equation}
 W = (Z_2)^N \ltimes \mathcal{S}_N
\end{equation}
And
 \bqa
 &\#W = 2^N N!\\
 &k = 2
 \eqa
\bqa
 I_b^{MNS}(Sp(2N)) = \f 2 {2^N N!} \f 1 {(2\pi i \e)^N} \int d^N u \prod_{i<j}
 \f {(u_i+u_j)^2(u_i-u_j)^2}  {((u_i+u_j)^2-\e^2)((u_i-u_j)^2-\e^2)}
     \prod_{i} \f { 4 u_i^2} { (4 u_i^2-\e^2) }
 \eqa

\subsection{ $SO(2N)$ or $D_N$}
The root system is the set
\begin{equation}
\{\a u\}=\{\pm u_i \pm u_j, \quad i<j\}
\end{equation}
The Weyl group is the subgroup of the $B_N$ Weyl group. It
consists of permutations and changing {\it even number} of signs
of $u_i$.
\begin{equation}
 W = Z_2^{N-1} \ltimes \mathcal{S}_N
\end{equation}
And
 \bqa
 &\#W = 2^{N-1} N!\\
 &k = 2
 \eqa
\bqa
 I_b^{MNS}(SO(2N)) = \f 2 {2^{N-1} N!} \f 1 {(2\pi i \e)^N} \int d^N u \prod_{i<j}
 \f {(u_i+u_j)^2(u_i-u_j)^2}  {((u_i+u_j)^2-\e^2)((u_i-u_j)^2-\e^2)}
 \eqa

\subsection{$G_2$}
The root system is the set
\begin{equation}
\{\a u\}=\{\f 1 {\sqrt{3}} (u_i - u_j), \quad i \neq j\} \cup
\{\pm \f 1 {\sqrt{3}} (u_i + u_j - 2u_k), \quad i \neq j \neq
k\},\qquad i,j,k \in \{1,2,3\}
\end{equation}
Note, that the root lattice is restricted to the plane
$\a_1+\a_2+\a_3=0$ and thus it is 2-dimensional.

The Weyl group consists of permutations and changing
simultaneously all signs of $u_i$.
\begin{equation}
 W = Z_2 \ltimes \mathcal{S}_3
\end{equation}
And
 \bqa
 &\#W = 12\\
 &k = 1
 \eqa
\begin{equation}
\bal I_b^{MNS}(G_2) =  \left .  \f {1} {12} \int du_1 du_2 \, \f 1
{(2\pi i \e)^2} \prod _{\a \in \Delta(G_2)} \f {\a u } {\a u +\e}
\right |_{u_3=-u_2-u_1} \eal
\end{equation}

\subsection{$F_4$}
The root system is the set
\begin{equation}
\{\a u\}=\{\pm u_i \} \cup \{ \pm u_i \pm u_j, \quad i \neq j\}
\cup \{\f 1  2 (\pm u_1 \pm u_2 \pm u_3 \pm u_4)\}
\end{equation}

The Weyl group is a group of automorphisms of the lattice $Q(D_4)$

\begin{equation}
 \end{equation}
And
 \bqa
 &\#W = 24*48=1152\\
 &k = 1/2
 \eqa
\begin{equation}
\bal I_b = \f {1} {2 \cdot 1152} \int d^4 u \, \f 1 {(2\pi i
\e)^r} \prod _{\a \in \Delta(F_4)} \f {\a u } {\a u +\e} \eal
\end{equation}

\subsection{$E_6$}
The root system is the set
\begin{equation}
\{\a u\}=\{\pm \sqrt{2 }u_7 \} \cup \{ u_i - u_j,  i,j \leq 6\}
\cup \{\f 1  2 ( \sum_{i=1}^{6} \e_i u_i \pm \sqrt{2} u_7 ),
\e_i=\pm 1 , \sum_{i=1}^{6} \e_i = 0 \}
\end{equation}
and the root lattice is restricted to the plane $\{a |
\sum_{i=1}^{6} a_i = 0 \}$.

The Weyl group is a group of automorphisms of the lattice $Q(E_6)$

And
 \bqa
 &\#W = 6!*72\\
 & k = 3\sqrt{2}
 \eqa
\begin{equation}
\bal I_b^{MNS}(E_6) = \left . \f {3 \sqrt{2}} {6!*72} \int du_2
\dots du_7 \, \f 1 {(2\pi i \e)^r} \prod _{\a \in \Delta(E_6)} \f
{\a u } {\a u +\e} \right |_{u_1=-(u_2+\dots+u_6)} \eal
\end{equation}

\section{Comparison of $I_d^{KS}$ and $I_b^{MNS}$ for $G_2$ group}

In the $G_2$ case the valid result $151/864$ obtained by
evaluating contour integral can be expressed in the manner similar
to (\ref{KSmod}),(\ref{so2nimp}).

The $G_2$ Weyl group contains 12 elements $w \in W$. The following
ones have non zero $\det(1-w)$:
 $2$ terms with $\det(1-w)=1$, $2$ terms with $\det(1-w)=3$ and $1$ term with $\det(1-w)=4$.
According to $\cite{KS}$ we should get
\begin{equation}
I_d^{KS}(G_2)=\f 1 {12} \lb 2\cdot 1 +2 \cdot \f 1 3 + \f 1 4 \rb
= \f {1} {12} \f {35} {12}
\end{equation}
but $I_b^{MNS}$ seems to be obtained as
\begin{equation}
I_b^{MNS}(G_2)= \f 1 {12} \lb 2 \cdot (1)^2 + 2 \cdot (\f 1 3)^2 +
(-2) \cdot (\f 1 4)^2 \rb = \f {1} {12} \f {151} {72}
\end{equation}

\end{document}